\documentclass[structabstract,twocolumn]{aa}  
\usepackage{natbib}
\usepackage{graphicx}
\usepackage{txfonts}

\usepackage{amssymb}
\usepackage[notref,notcite]{} %showkeys
\usepackage{color}

\def\lesssim{\buildrel < \over {_{\sim}}}

\newcommand\U[1]{{\,\rm #1}}
\newcommand\kms{km\,s^{-1}}
\newcommand\cmc{cm^{-3}}

\newcommand\rs[1]{_\mathrm{#1}}

\newcommand\RSNR{R\rs{SNR}}

\newcommand\ESN{E\rs{SN}}
\newcommand\Mej{M\rs{ej}}

\newcommand\Msun{M_\odot}
\newcommand\Vsh{V\rs{sh}}
\newcommand\Vej{V\rs{ej}}
\newcommand\vx{v\rs{x}}

\newcommand\TeovTp{T\rs{e}/T\rs{p}}
\newcommand\IbovIn{I\rs{b}/I\rs{n}}
\newcommand\betad{\beta\rs{down}}
\newcommand\epsCR{\epsilon\rs{CR}}
\newcommand\PCR{P\rs{CR}}

\newcommand\rhoztot{\rho\rs{0,tot}}
\newcommand\rhozion{\rho\rs{0,ion}}
\newcommand\ntot{n\rs{tot}}

\newcommand\hN{h\rs{N}}
\newcommand\pmax{p\rs{max}}
\newcommand\xiinj{\xi\rs{inj}}
\newcommand\etaTH{\eta\rs{TH}}

\begin{document}

\title{Cosmic Ray acceleration and Balmer emission from SNR~0509-67.5}

\author{G. Morlino\inst{1}\fnmsep\thanks{email: morlino@arcetri.astro.it}, 
            P. Blasi\inst{1,2}, R. Bandiera\inst{1} \and E. Amato\inst{1}
            }
\institute{$^1$INAF/Osservatorio Astrofisico di Arcetri, Largo E. Fermi, 5, 50125, Firenze, Italy \\
	      $^2$INFN/Gran Sasso Science Institute, viale F. Crispi 7, 67100 L'Aquila, Italy
               }

\date{Received 28 June 2013, 2013; accepted August 2013}

% \abstract{}{}{}{}{} 
% 5 {} token are mandatory
 
\abstract
  % context heading (optional), leave it empty if necessary  
{Observation of Balmer lines from the region around the forward shock of supernova remnants may provide precious information on the shock dynamics and on the efficiency of particle acceleration at the shock.}
{We calculate the Balmer line emission and the shape of the broad Balmer line for parameter values suitable for SNR~0509-67.5, as a function of the cosmic ray acceleration efficiency and of the level of thermal equilibration between electrons and protons behind the shock. This calculation aims at using the width of the broad Balmer line emission to infer the cosmic ray acceleration efficiency in this remnant.}
{We use the recently developed non-linear theory of diffusive shock acceleration in the presence of neutrals. The semi-analytical approach that we developed includes a description of magnetic field amplification as due to resonant streaming instability, the dynamical reaction of both accelerated particles and turbulent magnetic field on the shock, and all channels of interaction between neutral atoms and background plasma that change the shock dynamics.} 
{We achieve a quantitative assessment of the CR acceleration efficiency in SNR~0509-67.5 as a function of the shock velocity and different levels of electron-proton thermalization in the shock region. If the shock moves faster than $\sim 4500\U{\kms}$, one can conclude that particle acceleration must be taking place with efficiency of several tens of percent. For lower shock velocity the evidence for particle acceleration becomes less clear because of the uncertainty in the electron-ion equilibration downstream. We also discuss the role of future measurements of the narrow Balmer line.}
   {}
\keywords{acceleration of particles -- cosmic rays -- Balmer emission --
          SNR:0509-67.5 }

\maketitle

%________________________________________________________________

\section{Introduction}

Supernova remnants (SNRs) are the most likely sources of the bulk of Galactic cosmic rays (CRs). Gamma ray observations carried out with both space-borne telescopes (AGILE, FERMI-LAT) and Cherenkov telescopes (MAGIC, HESS, VERITAS) are collecting a growing body of evidence of hadrons accelerated in SNRs, although fundamental questions remain yet to be answered: direct evidence for CR acceleration up to the knee (in protons) is still missing, and the spectra of CRs as inferred from multifrequency observations appear to be somewhat softer than the basic predictions of the theory of diffusive shock acceleration (DSA). The Tycho SNR is currently the only instance of a young SNR in which the combined radio, X-ray and gamma ray observations appear to be unambiguously explained as the result of acceleration of protons up to an energy close to the ``knee" \cite[]{morlino}. 

Particle acceleration at SNR shocks is described by the non-linear theory of diffusive shock acceleration (NLDSA) \cite[see][for a review]{maldrury}. The formalism introduced by \cite{AmatoBlasi05,AmatoBlasi06} and \cite{Caprioli08} allows one to take into account the dynamical reaction of accelerated particles on the shock and the magnetic field amplification as due to streaming instability induced by the same accelerated particles.  Conservation of momentum and energy leads to the straightforward prediction that, in the presence of efficient particle acceleration, the temperature of the plasma downstream of the shock should be lower than in the case in which no acceleration takes place. Moreover, the dynamical reaction of accelerated particles on the background plasma induces the formation of a precursor upstream of the subshock. 

The acceleration efficiency typically required in order for SNRs to be the sources of CRs is of order $\sim 10\%$. In the pioneering paper by \cite{Chevalier78} it was first suggested that if the supernova explosion occurs in a partially ionized medium, the observation of the Balmer line width in proximity of the collisionless shock that develops can provide information on the temperature of the plasma and therefore on the efficiency of CR acceleration. More specifically, the broad Balmer line is expected to be narrower, reflecting a lower proton temperature downstream, and the narrow Balmer line is expected to be broader, reflecting heating in the precursor. 

Despite the simple expectations illustrated above, a comprehensive description of the behavior of a collisionless supernova shock in the presence of neutrals, for arbitrary shock speed, was only recently put forward. A first effort to include the presence of neutrals in the shock acceleration theory was done by \cite{Wagner09} using a two-fluid model to treat ions and CRs but neglecting the dynamical role of neutrals. A different model was proposed by \cite{Raymond11}: here momentum and energy transfer between ions and neutrals is included, but both the CR spectrum and the profile of the CR-precursor are assumed {\it a priori} rather than calculated self-consistently.
\cite{paperI} proposed a semi-analytical kinetic calculation that returns the distribution function of neutral hydrogen atoms in the absence of accelerated particles, and all thermodynamical properties of the ion plasma. In that paper it was shown that a substantial fraction of neutral atoms that suffer a charge exchange (CE) reaction downstream produce neutral atoms that move towards upstream and can release energy and momentum in the upstream plasma, thereby heating it. \cite{paperII} showed that this phenomenon of {\it neutral return flux} creates an intermediate component of the Balmer line. Both these calculations were later used by \cite{paperIII} to generalize the theory of NLDSA to include the effect of neutrals. This theory allows one to calculate the spectra of accelerated particles, the magnetic field amplification at the shock and the modifications induced on the shock by neutral atoms. The processes of CE and ionization, as well as excitation of atoms as due to collisions with both ions and electrons, are included in the calculations: all the elements are provided to accurately calculate the Balmer emission, as discussed in the article by \cite{paperII}. 

Application of this detailed modeling to individual SNRs in which the strength and shape of the Balmer line are accurately measured may finally provide quantitative information on the efficiency of particle acceleration in these sources.
In SNR~0509-67.5, a Balmer-dominated shock was discovered already in the 80's \citep{Tuohy82}. Initially no broad Balmer line was detected because of the very low flux level, and detection of the broad component only arrived with the work by \cite{Ghava07}. These authors measured the width of the broad component of the Ly$\beta$ line to be $3700\pm400\U{\kms}$. However, as the spectrum was taken from the entire remnant, it remains uncertain whether the obtained line width is broadened by the bulk motion of the plasma. Only recently \cite{Helder10} were able to carry out a measurement of the broad component of the H$\alpha$ line emission in two different regions of the blast wave, located in the southwest (SW) and northeast (NE) rim, obtaining a FWHM of $2680\pm 70\U{\kms}$ and $3900 \pm 800\U{\kms}$, respectively. The shock velocity was estimated to be $\Vsh=6000\pm 300\U{\kms}$ when averaged over the entire remnant, and $6600\pm 400\U{\kms}$ in the NE part. The width of the broad Balmer line was claimed by the authors to be suggestive of efficient CR acceleration.

The most important sources of uncertainty in deriving the CR acceleration efficiency for this SNR (and also in general) derive from the poor knowledge of the bulk speed of the forward shock and of the level of electron-ion thermalization behind the shock front. In order to extract information on CR acceleration from the measurement of the Balmer line, a theory is needed that accounts for the many subtleties of the modifications induced by both CRs and neutral atoms on the structure of a collisionless shock. Here we use the theoretical framework of \cite{paperIII} to give a quantitative assessment of the evidence for CR acceleration in SNR~0509-67.5.

The paper structure is as follows: in \S~\ref{sec:dynamics} we discuss the existing estimates of the shock velocity and how these estimates compare with the results of a simple model of the dynamics of this remnant. The results of our calculations, including the dynamical reaction of accelerated particles and the shock modification induced by neutrals, are illustrated in \S~\ref{sec:results}. In \S~\ref{sec:narrow} we briefly discuss some predictions on the width of the narrow Balmer line in SNW~0509-67.5, with and without particle acceleration. We summarize in \S~\ref{sec:conclusion}.

\section{Remnant dynamics}
\label{sec:dynamics}

At present, we are not aware of any direct measurement of the proper motion of the shock in SNR~0509-67.5. \cite{Helder10} estimated the shock velocity by comparing the measured width of X-ray lines produced by shocked ejecta with the expectation of an evolutionary model of both the forward and reverse shock. These authors obtain for the entire remnant $\Vsh= 6000\pm300\U{\kms}$, while they give an estimate of $5000\U{\kms}$ for the SW region. 

The shock speed can also be estimated from measurements of the SNR age and shock radius. SNR~0509-67.5 is located in the LMC, so that its distance from the Sun is known to be $50\pm1\U{kpc}$. The age estimated from the light echo is $400\pm120\U{yr}$ \citep{Rest05}, although the spectral and dynamical properties of the SNR, together with some historical considerations, constrain the value to $400\pm50\U{yr}$ \citep{Badenes08}. This remnant originated from a type Ia SN, probably produced by a double-degenerate system \citep{Schaefer12}. Based on the high quality of the light echo, \cite{Rest07} concluded that the parent event of SNR~0509-67.5 belonged to the group of overluminous, highly energetic type Ia SNe whose prototype is SN 1991Ts. Interestingly enough the same conclusion has been reached by \cite{Badenes08} based on the X-ray spectrum: their best model has a kinetic energy of $1.4\times 10^{51}$ erg and a mass density of the circumstellar medium $\rho_0= 10^{-24}\U{\cmc}$. 
    
At first sight the dynamics of~0509-67.5 may be constrained by two pieces of observational evidence. The first is the angular radius measured by Chandra in the X-ray band \cite[$14.8''\pm0.05''$;][]{Warren04}, which translates into a linear radius of $1.1 \times 10^{19}\U{cm}$ ($3.6\U{pc}$) at the known distance of the LMC. These measurements can be considered very accurate, because the errors on both angular radius and distance are only at the few percent level. Moreover \cite{Badenes07} concluded that the expansion does not occur in a cavity, as would be expected for a type II SN explosion, but rather in the normal ISM.

On the other hand, a more careful look at the morphology of this SNR shows that its shape deviates appreciably from spherical. The NE limb looks remarkably circular \cite[see e.g. the H$\alpha$ image in the paper by][]{Helder10}. By fitting a circle to the portion of the limb with polar angle ranging from 20 to 140~degrees, we derive a mean radius of $16.1\U{arcsec}$ (namely $3.9\U{pc}$) on that side, with average deviations less than $0.07\U{arcsec}$ from that curve. However, assuming that the center of this circle coincides with the physical location of the supernova explosion, the distance to the SW limb can be estimated to be $13.2\U{arcsec}$ (corresponding to $3.2\U{pc}$).

Even though these estimates depend on the conjecture that deviations from the spherical symmetry are more prominent on the SW side, this appears at least qualitatively reasonable given the strong enhancement on that side of dust emission, seen both with Spitzer \cite[]{Borkowski06} and with AKARI \cite[]{Seok08}, also accompanied by brighter X-ray limb emission \cite[]{Warren04} and Balmer emission. Moreover the H$\alpha$ line emission in the SW limb shows a higher value of the ratio $\IbovIn$ compared with the NE region, compatible with a lower shock velocity.

\begin{figure}
\begin{center}
{\includegraphics[width=\linewidth]{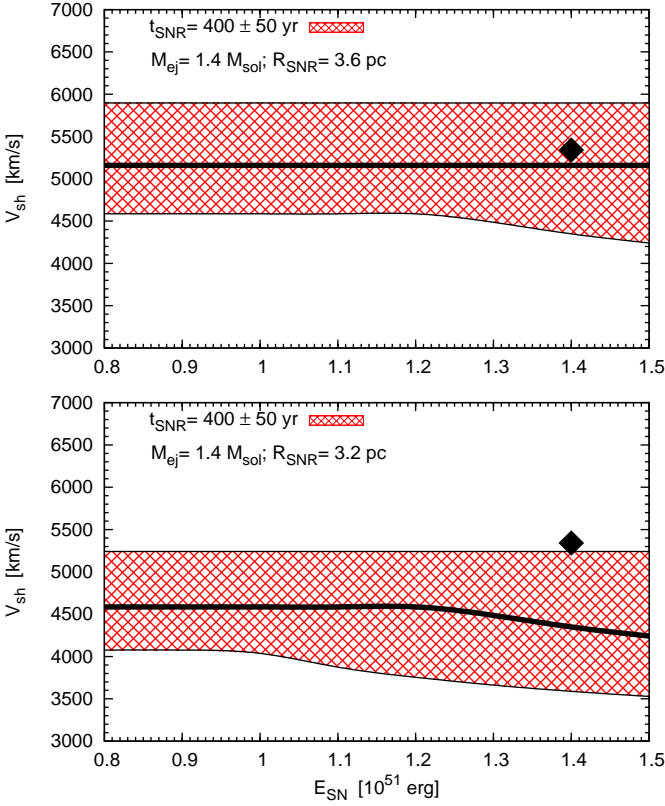}}
  \caption{Velocity of the forward shock as a function of explosion energy calculated using the theoretical model of \cite{Truelove99} for a shock expanding into a uniform medium with density $\rho_0$, and assuming a power law density profile for the ejecta with power law index $n=7$. The upper (lower) panel refers to a SNR radius $\RSNR=3.6\U{pc}$ ($\RSNR=3.2\U{pc}$). The thick curve is obtained matching the  observational constraints derived from the remnant radius ($3.6\U{pc}$ for the upper panel and $3.2\U{pc}$ for the lower panel) and age ($400\pm50\U{yr}$). The mass of the ejecta is fixed at $\Mej=1.4\,\Msun$. The hatched area shows the results allowing for the uncertainty in the age of the remnant (the lower side corresponding to $450\U{yr}$ while the upper to $350\U{yr}$).  The filled diamond shows the shock velocity calculated using the the best fit parameters obtained by \cite{Badenes08}, i.e. $\ESN=1.4 \times 10^{51}$ erg and $\rho_0=10^{-24}\U{g\,\cmc}$.}
  \label{fig:velocity}
\end{center}
\end{figure}

Here we try to identify the range of meaningful values of the shock velocity in the two regions by modelling the dynamics of SNR0509-67.5 according to the analytical description of \cite{Truelove99} for the evolution of a remnant expanding into a uniform medium with density $\rho_0$. Following \cite{Badenes08} we fix the mass of the ejecta to the standard value $\Mej=1.4\,\Msun$. The structure function of the ejecta is taken as $\propto (v/\Vej)^{-7}$ \cite[see \S~3.2 and 9 in][]{Truelove99}. The explosion energy $\ESN$ is taken as a free parameter. For each value of $\ESN$ there is only one value of $\rho_0$ for which the calculated radius matches the observed value. In Fig.~\ref{fig:velocity} we plot the shock velocity as a function of the SN energy for a SNR radius of $3.6\U{pc}$ (the average over the whole remnant) in the upper panel and $3.2\U{pc}$, probably more appropriate to the SW rim, in the lower panel. In both panels the hatched area illustrates the uncertainty in the age of the remnant. The diamond symbol represents the shock velocity calculated using the the best fit parameters obtained by \cite{Badenes08}, i.e. $\ESN=1.4 \times 10^{51}$ erg and $\rho_0=10^{-24}\U{\cmc}$, which give $\Vsh= 5340\U{\kms}$. One should however keep in mind that this estimate was obtained for the whole remnant, while the relevant velocity for our calculations is the local shock speed in the region where the Balmer emission is measured. In fact, for the SW rim ($\RSNR=3.2\U{pc}$) one can see that shock velocities as low as $\sim 3500\U{\kms}$ are still marginally compatible with the dynamics of this SNR. 

In order to take into account the uncertainty in $\Vsh$, reflecting the different results found by  \cite{Badenes08}, \cite{Helder10} and our own calculations, we carry out the calculation of Balmer emission for $\Vsh=4000$ and $5000\U{\kms}$ for the SW rim and for $6000\U{\kms}$ for the NE rim.

Calculations of the Balmer emission are also affected rather heavily by uncertainties in the electron-ion thermalization downstream of the shock. For the values of gas density typical of the ISM, the SNR shock is collisionless and electrons and protons are not expected to equilibrate rapidly to the same temperature. While Coulomb scattering is usually far too slow to bring electrons and protons in thermal equilibrium, some collective effects may lead to at least some partial equilibration, though our poor knowledge of the processes at work makes it impossible to assess the issue quantitatively. Below, we use the ratio $\betad$ of electron to proton temperature as a parameter, while phenomenological hints on its value can be found in the literature.

By modeling the X-ray spectra of shocked ejecta, \cite{Badenes08} suggest that the allowed range of values for $\betad$ is $0.01<\betad<0.1$. Although this result strictly applies to the reverse shock, for this SNR the speed of the latter is expected to be not much lower than that of the forward shock, based on evolutionary models. Therefore a similar range could be expected for the outer shock. If so, electrons and protons appear to be rather far from thermal equilibrium. 

As a more general statement, the current literature, based on various techniques, seems to be converging on a consensus that the level of equilibration between electrons and protons becomes lower with increasing shock velocity \cite[see][for a review]{Rakowski05}. As a conservative estimate, for $\Vsh>3000\U{\kms}$ one can take $\TeovTp\lesssim0.2$ \cite[see fig.~6 in][]{Rakowski05}. A comprehensive review of the problem of electron-ion equilibration in collisionless shocks and the role of Balmer line measurements has recently appeared in literature \cite[]{ghavarev}.

Special caution is needed in using these estimates to infer the efficiency of particle acceleration., because some of them are derived by making use of the measured intensity ratio of broad to narrow Balmer lines. This quantity is actually also sensitive to the efficiency of particle acceleration, so that the information it contains becomes degenerate when this process is important.

\section{FWHM of the broad Balmer line and CR acceleration}
\label{sec:results}

\begin{figure}
\begin{center}
{\includegraphics[width=\linewidth]{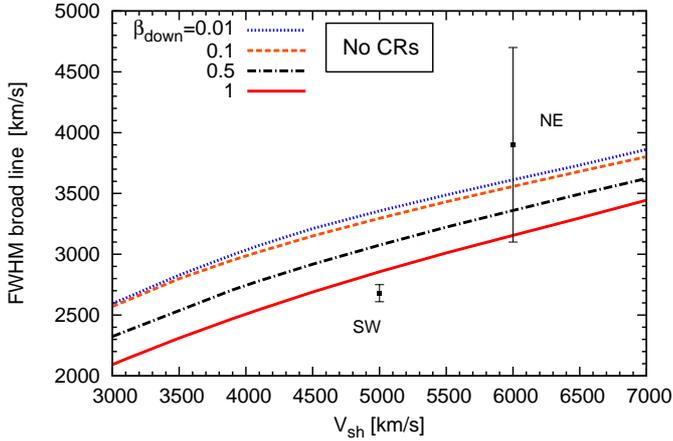}}
\caption{FWHM of the broad line in the absence of CRs and for different levels of electron-ion equilibration. The two data points represent the FWHM as measured by {\protect \cite{Helder10}} in the NE and SW regions of SNR~0509-67.5. Because the shock velocity is rather uncertain, the horizontal coordinate of the two points can actually be appreciably different from that in the figure, which is taken directly from the work of {\protect \cite{Helder10}}.}
\label{fig:FWHM_broad}
\end{center}
\end{figure}

\begin{figure*}
\begin{center}
{\includegraphics[width=1\linewidth]{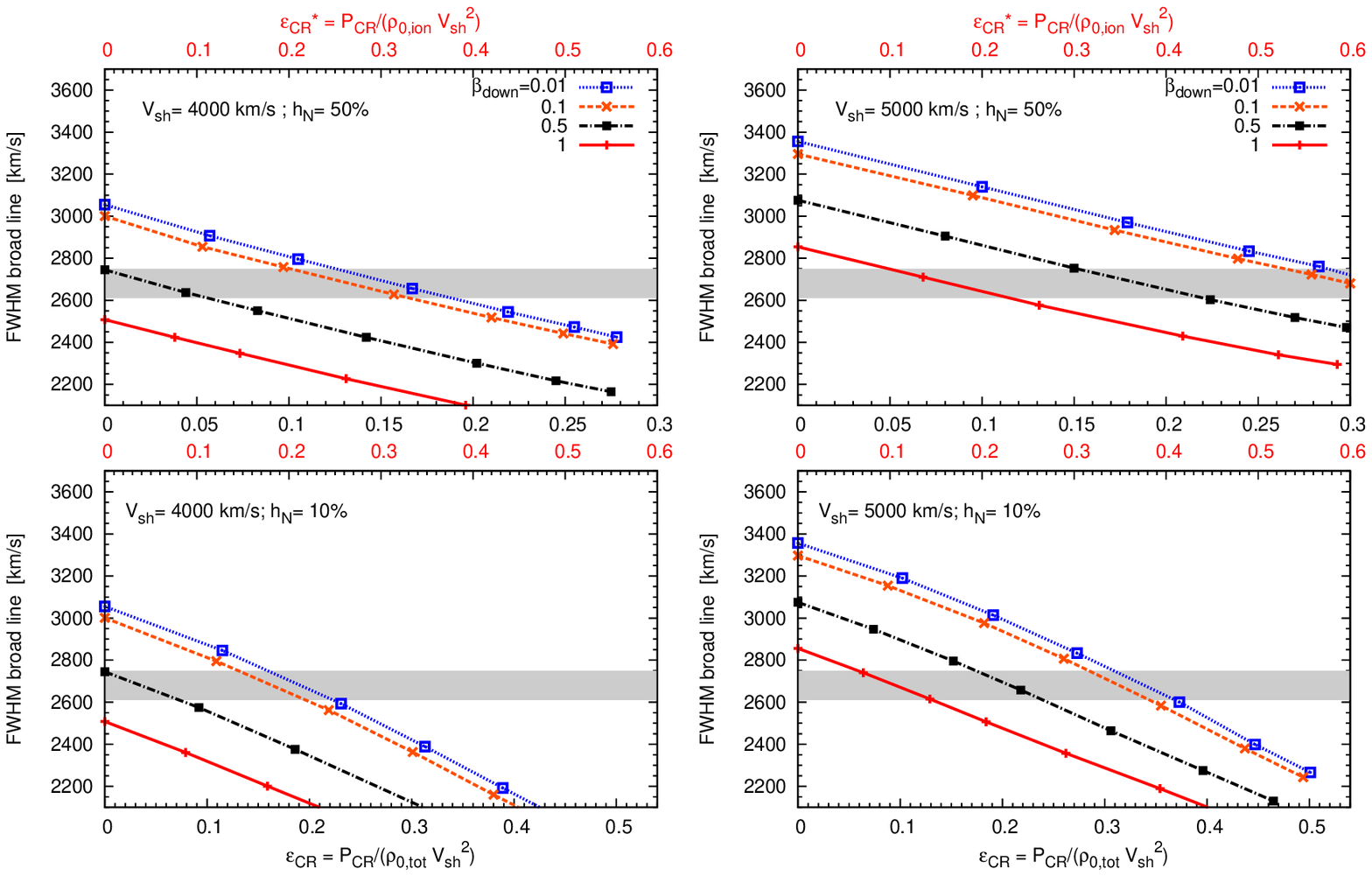}}
\caption{FWHM of the broad Balmer line as a function of the CR acceleration efficiency. Different lines show different electron-ion equilibration levels. The plots on the left (right) are obtained for $\Vsh=4000\U{\kms}$ ($\Vsh=5000\U{\kms}$). The top (bottom) panels refer to neutral fraction $\hN=50\%$ ($\hN=10\%$). The total gas density is $n_0=0.6\U{\cmc}$ \cite[]{Badenes08}. The shaded area shows the FWHM as measured for the SW shock by {\protect \cite{Helder10}} at 1 $\sigma$ level.}
\label{fig:FWHM_broad_CR-V4+5}
\end{center}
\end{figure*}

The calculations of the shock structure of SNR~0509-67.5 are carried out by using the formalism of \cite{paperIII}, where both the CR back-reaction and the shock modification induced by neutral hydrogen are taken into account. As discussed by \cite{paperIV}, the kinetic approach adopted for the description of the neutral distribution function is crucial in this type of calculations: for fast shocks ($\Vsh>2500\U{\kms}$) neutral hydrogen atoms do not reach thermal equilibrium with the background ions despite the CE reactions that couple them together. This is due to the fact that ionization occurs before such equilibration can be reached. As a consequence, the width of the Balmer line, that reflects the mean energy per particle of neutrals downstream of the shock, is somewhat smaller than predicted by assuming thermalization after a few CE reactions, as was done for instance by \cite{vanAdel08}. Since our kinetic calculations follow the evolution of the distribution function of neutral hydrogen in detail, we are able to predict the shape of the Balmer line very reliably, without any assumption on it being a Maxwellian.

In Fig.~\ref{fig:FWHM_broad} we show the FWHM of the broad line as a function of shock speed, as obtained from our calculations in the absence of CR acceleration. The two data points represent the FWHM as measured by \cite{Helder10} in the NE and SW regions of SNR~0509-67.5. As we discussed above, the shock velocity is rather uncertain, which implies that the horizontal coordinate of the two points can actually be appreciably different from that in the figure, which is taken directly from the work of \cite{Helder10}. The solid (dotted) line is the FWHM we predict in the absence of CR acceleration and for $\betad=1$ ($\betad=0.01$). The large error bar on the FWHM in the NE rim makes it difficult to infer anything about particle acceleration in that region of the shock. On the other hand, in the SW region, the measured FWHM of the broad Balmer line is appreciably below the prediction, whatever the level of electron-proton thermalization, if the shock velocity is higher than $\sim 4500\U{\kms}$. This fact strongly suggests that CR acceleration may be taking place in the SW region of the remnant. 

The calculations are therefore repeated for the SW rim, for two values of the shock velocity and varying the CR acceleration efficiency. The CR acceleration efficiency is defined here as $\epsCR=\PCR/\rhoztot\Vsh^{2}$ although another definition is more meaningful from the physical point of view: $\epsCR^{*}=\PCR/\rhozion\Vsh^{2}$, where $\rhozion= (1-\hN)\,\rhoztot$ is the density of ionized material at upstream infinity and $\hN$ is the neutral fraction. One should keep in mind that collisionless shocks only act on the ionized component of the background fluid, and that only ions can take part in the acceleration process. 

\begin{figure}
\begin{center}
{\includegraphics[width=\linewidth]{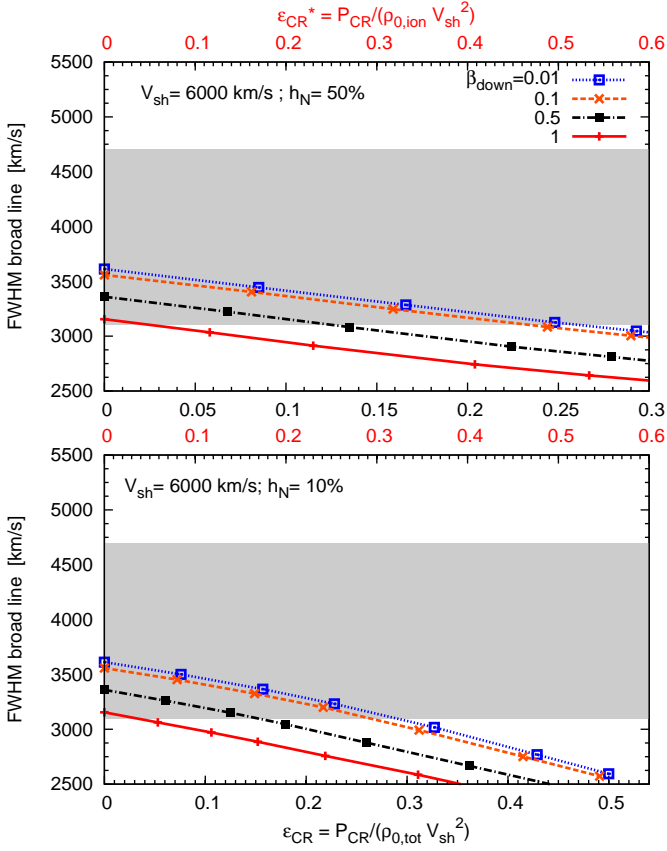}}
\caption{The same as in Fig.~\ref{fig:FWHM_broad_CR-V4+5} but for shock velocity $\Vsh= 6000\U{\kms}$. Here the shaded area shows the FWHM measured for the NE sector of the shock by {\protect \cite{Helder10}} at 1 $\sigma$ level.}
\label{fig:FWHM_broad_CR-V6}
\end{center}
\end{figure}

The FWHM of the broad Balmer line as a function of $\epsCR$ (on the lower $x$-axis) and $\epsCR^{*}$ (upper $x$-axis) is reported in Fig.~\ref{fig:FWHM_broad_CR-V4+5}. The fraction of neutrals far upstream is taken as $\hN=50\%$ in the upper panels and $\hN=10\%$ in the lower panels. The two columns correspond
to different values of the shock velocity: $\Vsh=4000\U{\kms}$ (on the left) and $\Vsh=5000\U{\kms}$ (on the right), which are expected to bracket the most likely interval of values of the shock velocity in the SW rim. The four curves in each panel refer to different levels of electron-ion thermalization, as specified in the figure. The shaded area in each panel shows the measured FWHM \cite[]{Helder10} with the $1\sigma$ error bar. 

As one expected, the FWHM of the broad Balmer line decreases with increasing acceleration efficiency. The intersection between each line and the shaded area identifies the CR acceleration efficiency required to explain the data. For $\Vsh=5000\U{\kms}$ and neutral fraction $\hN=50\%$, $\epsCR^{*}$ (top $x$-axis) ranges from $10 - 23\%$, for the case of full equilibration ($\betad=1$), to $>57\%$ for $\betad=0.01$. For the same value of the shock velocity and 10\% neutral fraction, $\epsCR^{*}$ ranges from $\sim 10\%$ for the case of full equilibration ($\betad=1$) to $\sim 35-40\%$ for $\betad=0.01$. As expected, the evidence for CR acceleration becomes weaker when the shock velocity is lower: for $\Vsh=4000\U{\kms}$ there is no evidence for particle acceleration if $\betad=1$ (full equilibration), for any value of the ionization fraction. On the other hand, if $\betad=0.1-0.01$, the measured FWHM is still compatible with $\epsCR^{*}\simeq 20-30\%$ for $\hN=50\%$ and $\sim 20\%$ for $\hN=10\%$.

The calculation presented in Fig.~\ref{fig:FWHM_broad_CR-V4+5} has been carried out assuming the best fit value for the upstream density inferred by \cite{Badenes08}, namely $n_0= 0.6\U{\cmc}$. It is worth pointing out that in the absence of CR acceleration the predicted FWHM does not depend on the value of the density; for efficient CR acceleration the FWHM changes very slightly by changing the gas density: lowering the density to $0.1\U{\cmc}$ the FWHM changes only at the level of $\sim 1\%$ for $\epsCR^{*} > 50\%$. For lower values of $\epsCR^{*}$ we find that the variation is totally negligible.

For the sake of completeness, we also calculated the Balmer emission expected from the NE rim, with $\Vsh=6000\U{\kms}$. Our results are shown in Fig.~\ref{fig:FWHM_broad_CR-V6} for $\hN=50\%$ (top panel) and  $\hN=10\%$ (lower panel). The shaded area illustrates the measured FWHM \cite[]{Helder10}. As expected, in this case we can only derive an upper limit on the acceleration efficiency: $\epsCR^{*} < 50\%$ for $\hN=50\%$ and $\epsCR^{*} < 30\%$ for $\hN=10\%$.

An estimate of the acceleration efficiency in the SW rim was also presented by \cite{Helder10}, based on calculations previously carried out by \cite{vanAdel08} and on generalized Rankine-Hugoniot relations including the pressure of accelerated particles and their escape. Such calculations of the width of the Balmer line were not based on self-consistent modeling of the non-linear interactions that determine the shock structure: the shock modification due to neutrals and accelerated particles was only described in a phenomenological way. Moreover, as discussed in \cite{paperIV}, the calculations presented by \cite{vanAdel08} are based on the assumption that the distribution function of neutral hydrogen atoms becomes Maxwellian, with the same temperature as for protons, for particles that have suffered more than two reactions of CE downstream. Our kinetic calculations show that this is not the case (see \cite{paperIV}, where the comparison with the results of \cite{vanAdel08} was discussed in detail). The FWHM derived by \cite{vanAdel08} is close to the results of our kinetic calculations only for shocks with $\Vsh < 2500\U{\kms}$, but is systematically larger than our findings for faster shocks. This is due to the fact that for fast shocks, ionization occurs before a large number of CE reactions can drive ions and neutrals towards thermal equilibrium, so that the assumptions made by \cite{vanAdel08} fail and a full kinetic calculation is required. By making a simple linear extrapolation of the curves in \cite{vanAdel08}, \cite{Helder10} derive a lower limit on the acceleration efficiency of $>15\%$, obtained for full electron-ion equilibration. Given the limitations of such a simple approach, this lower limit can be considered in reasonably good agreement with our more detailed determination of the acceleration efficiency, if the shock velocity is assumed to be $\Vsh=5000\U{\kms}$, as in \cite{Helder10}. Given the uncertainties in the shock velocity, this bound should not be taken as an absolute lower limit.

As discussed above, in the case when electrons and ions downstream are close to full equilibration ($\betad\sim 1$), we find that the data are compatible with the absence of particle acceleration if the shock velocity is below $\sim4500\U{\kms}$, a value that still fits well with models for the dynamical evolution of the SNR (see Fig.~\ref{fig:velocity}). The rapid electron-ion equilibration required, however, is somewhat at odds with the expectation for a fast collisionless shock, although it cannot be ruled out as yet. On the other hand, if the equilibration were inefficient ($\betad\lesssim 0.1$) then the measurement would be compatible with the absence of CR acceleration only for a shock speed $\lesssim 3500\U{\kms}$.

\section{The narrow line}
\label{sec:narrow}

A possible way to break the degeneracy between the acceleration efficiency and the electron-ion equilibration would be to measure the narrow Balmer line emission from the same region of the shock where the broad line has been observed. In the presence of effective particle acceleration a precursor is formed upstream of the shock that may lead to a broadening of the narrow line due to CE reactions between neutrals and ions inside the precursor \cite[see][for a discussion]{paperIII}. The combination of measurements of the widths of the two components can in principle allow a determination of the CR acceleration efficiency, although it is worth keeping in mind that additional parameters enter the calculation of the narrow line in the presence of CR acceleration: for instance, the maximum energy of accelerated particles and the level of turbulent heating in the precursor. From the observational point of view, \cite{Smith94} reported the measurement of the FWHM in three different regions, the center and east rim, both having a FWHM of $25\pm2\U{\kms}$, and the west rim, with a FWHM of $31\pm2\U{\kms}$. While the former are compatible with the standard width of the narrow Balmer line in the ISM, the latter is suggestive of a broadening that can be due to the presence of CRs. Moreover the shape of the line in the west rim shows evidence for non Gaussian wings, which can also result from the presence of accelerated particles \cite[and discussion below]{paperIII}. One should however keep in mind that these measurements refer to regions that are spatially different from those where the broad Balmer line has been measured. This is the reason why here we concentrated our attention on the detailed measurements of the shape and intensity of the broad component only. 

On the other hand, we think it useful to show how the shape of the narrow Balmer line changes depending on the efficiency of particle acceleration and of non-adiabatic heating in the CR precursor. We selected the case with $\Vsh= 5000\U{\kms}$, $\ntot= 0.6\U{\cmc}$, $\hN= 10\%$, $\betad= 0.5$ as the best suited to illustrate the effect of CRs on the narrow Balmer line. The maximum momentum and the injection parameter are fixed to $\pmax= 10\U{TeV}$ and $\xiinj= 3.7$ respectively, so as to have $\epsCR^* \simeq 25\%$, compatible with the FWHM of the broad Balmer line as shown in Fig.~\ref{fig:FWHM_broad_CR-V4+5}.

\begin{figure}
\begin{center}
{\includegraphics[width=1\linewidth]{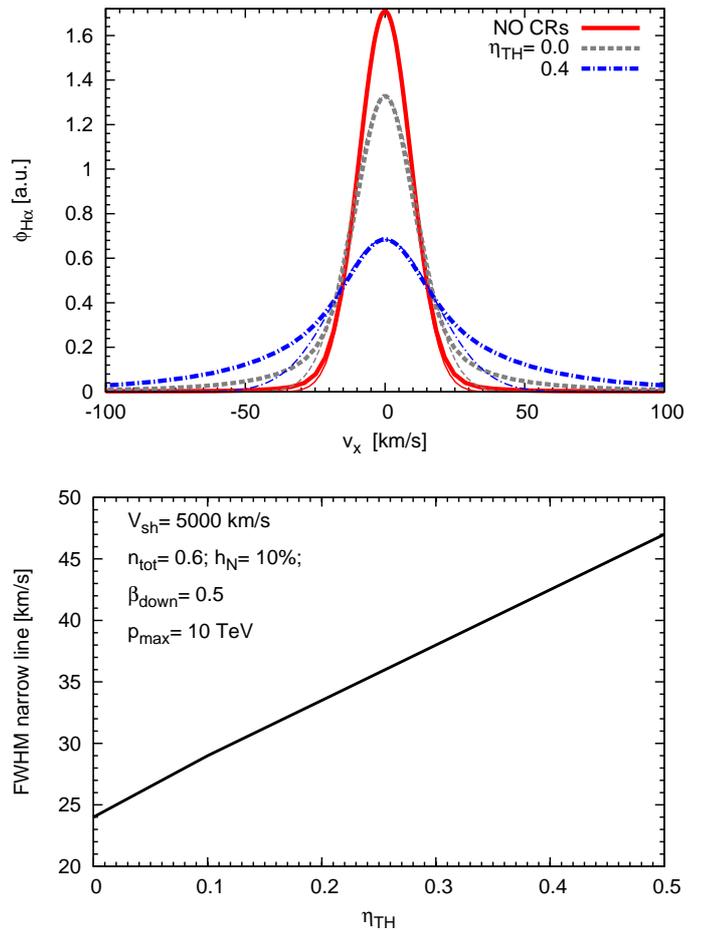}}
  \caption{{\it Top panel}: thick lines show the profile of the narrow Balmer line for three cases: no CR acceleration (solid line) and acceleration efficiency $\epsCR^{*}= 25\%$ (i.e. $\epsCR= 22\%$) for two different values of the turbulent heating parameter, $\etaTH=0$ (dashed) and 0.4 (dot-dashed). For all the three cases the thin lines show the corresponding Gaussian having the same FWHM of the narrow Balmer line.  {\it Bottom panel}: FWHM of the narrow Balmer line as a function of $\etaTH$, for acceleration efficiency $\epsCR^{*}= 25\%$. For both panels the values of the other parameters are fixed as shown in the label of the bottom panel.}
  \label{fig:FWHM_narrow}
\end{center}
\end{figure}

In the top panel of Fig.~\ref{fig:FWHM_narrow} we plot the predicted shape of the narrow Balmer line for two values (0 and 0.4) of the turbulent heating parameter $\etaTH$  \cite[namely the amount of turbulent magnetic field that is eventually converted to thermal energy of the plasma upstream; see][]{paperIII}. For comparison, we also show the case without CR acceleration (solid line). In the bottom panel we plot the FWHM of the narrow Balmer line as a function of $\etaTH$.   

Let us discuss this latter panel first. Here we show that the FWHM of the narrow line is very sensitive to the value of $\etaTH$. On the other hand, the effect produced by adiabatic compression is very small: in the absence of any non-adiabatic heating ($\etaTH=0$), the FWHM is  only $24$, rather than $21\U{\kms}$, which is the width corresponding to the upstream temperature of $10^4\U{K}$ in the absence of CR acceleration.

The line profile for the case when CR acceleration is absent (FWHM of $21\U{\kms}$) is shown as the thick solid line in the top panel of Fig.~\ref{fig:FWHM_narrow}. The thin solid line shows a gaussian shape with the same width. The weak wings that appear at high $\vx$ are the result of the neutral return flux discussed by \cite{paperI}: this phenomenon is very weak for shock velocity $>3000\U{\kms}$.
 
The dashed and dash-dotted lines in the top panel of Fig.~\ref{fig:FWHM_narrow} show the shape of the narrow Balmer line for $\etaTH=0$ and 0.4. One can appreciate that in both cases the width of the narrow Balmer line increases, more prominently for the larger value of $\etaTH$. In addition to the broadening however, one can also notice the rise of non-gaussian wings (differences between thick and thin lines of the same type), which are not due to the neutral return flux (relatively small for large shock velocities) but rather to the CR-induced precursor upstream. A broadening of the narrow Balmer line and the appearance of non-gaussian wings are the imprint of CR acceleration. 

\section{Conclusions}
\label{sec:conclusion}

The formalism proposed by \cite{paperI} and \cite{paperII,paperIII} to describe the process of diffusive particle acceleration at shocks in the presence of neutral hydrogen atoms was applied here to SNR~0509-67.5, for which measurements of the width of the broad Balmer line have recently become available \cite[]{Helder10}. Such measurements were carried out on the SW and NE rim of the remnant, returning a FWHM of $2680\pm 70\U{\kms}$ and $3900 \pm 800\U{\kms}$, respectively.

For the NE rim, the FWHM of the broad line is fully consistent with standard shock dynamics in the absence of accelerated particles, for a shock velocity of $6000\U{\kms}$. For the SW rim, \cite{Helder10} use $5000\U{\kms}$ as a lower limit to the shock velocity, based on a simple model of the dynamics of the remnant. On the other hand, we showed that the asymmetry between the NE and SW parts of the SNR suggest a somewhat lower velocity for the SW rim. Our calculations show that shock velocities between 3500 and $5500\U{\kms}$ are still compatible with the measured angular size of the remnant in the SW region. In the NE, a shock velocity of $\sim 6000\U{\kms}$ seems appropriate. 

In order to quantify the impact of the uncertainty in the shock velocity on the estimate of the CR acceleration efficiency, we carried out the calculations of the shock dynamics, of particle acceleration and of Balmer line emission for $\Vsh=4000$ and $5000\U{\kms}$ in the case of the SW rim and for $\Vsh=6000\U{\kms}$ for the NE rim. For both rims two values of the neutral fraction at upstream infinity were considered: $\hN=50\%$ and $\hN=10\%$.

If the shock speed in the SW region is $5000\U{\kms}$, our calculations show evidence for particle acceleration with efficiency between $\sim 10\%$ and $\sim 50\%$, depending on the level of thermal equilibration between electrons and ions downstream of the shock. The inferred CR acceleration efficiency depends very weakly upon the value of the gas density around the SNR. Lower shock speeds lead to a less clear situation: for $\Vsh=4000\U{\kms}$, the evidence of CR acceleration appears to be solid only if a low level of thermal equilibration between ions and electrons is assumed ($\betad\sim 0.1-0.01$), in which case the CR acceleration efficiency is at the level of $\sim 20-30\%$. 

In the NE rim, the error bar on the measured value of the FWHM is so large that it is not possible to infer any firm conclusion in terms of particle acceleration. 

Measurements of the width of the narrow Balmer line at the same spatial locations where the broad Balmer emission is measured may help resolve the degeneracy between CR acceleration and electron-ion equilibration, but such measurements are so far not available for this SNR. However, we also calculated the shape of the narrow Balmer line for the same set of parameters that appear to describe the broad emission, so as to show the predictive power associated with the measurement of the narrow component. We find that the presence of CRs certainly leads to a broadening of the narrow Balmer line, but this effect is rather sensitive to the fraction $\etaTH$ of turbulent field energy density that is transformed into heating of the upstream plasma. In addition to a larger FWHM of the narrow line (typical widths range between 30 and $50\U{\kms}$, to be compared with $21\U{\kms}$ expected in the absence of CR acceleration), the presence of a CR-induced precursor leads to the appearance of non-gaussian wings in the shape of the line, more pronounced for larger values of $\etaTH$. 

\begin{acknowledgements}
We are grateful to D. Caprioli for discussions on the topic. 
This work was partially funded through grant PRIN INAF 2010 and ASTRI. 
\end{acknowledgements}

\end{document}